\documentclass[twoside,twocolumn,aps,prc,superscriptaddress,nofootinbib,showpacs]{revtex4}
\usepackage{amsmath,amssymb,graphicx}

\allowdisplaybreaks

\begin{document}

\title{$\Sigma_c {\bar D}$ and $\Lambda_c {\bar D}$ states in a chiral quark model}

\author{W.L.~Wang}
 \affiliation{Institute of High Energy Physics, CAS, P.O. Box 918-4, Beijing 100049, China}
 \affiliation{Theoretical Physics Center for Science Facilities (TPCSF), CAS, Beijing 100049, China}

\author{F.~Huang}
 \affiliation{Department of Physics and Astronomy, The University of Georgia, Athens, GA 30602, USA}

\author{Z.Y.~Zhang}
 \affiliation{Institute of High Energy Physics, CAS, P.O. Box 918-4, Beijing 100049, China}
 \affiliation{Theoretical Physics Center for Science Facilities (TPCSF), CAS, Beijing 100049, China}

\author{B.S.~Zou}
 \affiliation{Institute of High Energy Physics, CAS, P.O. Box 918-4, Beijing 100049, China}
 \affiliation{Theoretical Physics Center for Science Facilities (TPCSF), CAS, Beijing 100049, China}

\begin{abstract}
The $S$-wave $\Sigma_c \bar D$ and $\Lambda_c \bar D$ states with
isospin $I=1/2$ and spin $S=1/2$ are dynamically investigated within
the framework of a chiral constituent quark model by solving a
resonating group method (RGM) equation. The results show that the
interaction between $\Sigma_c$ and $\bar D$ is attractive, which
consequently results in a $\Sigma_c \bar D$ bound state with the
binding energy of about $5-42 $ MeV, unlike the case of $\Lambda_c
\bar D$ state, which has a repulsive interaction and thus is
unbound. The channel coupling effect of $\Sigma_c \bar D$ and
$\Lambda_c \bar D$ is found to be negligible due to the fact that
the gap between the $\Sigma_c \bar D$ and $\Lambda_c \bar D$
thresholds is relatively large and the $\Sigma_c \bar D$ and
$\Lambda_c \bar D$ transition interaction is weak.
\end{abstract}

\pacs{13.75.Jz, 12.39.-x, 14.20.Gk}

\keywords{$\Sigma_c \bar D$ and $\Lambda_c \bar D$ states; Quark model}

\maketitle

\section{Introduction}

Understanding the structure and dynamical origin of baryon
resonances is one of the most important topics within the field of
hadron physics. On quark level, several constituent quark models
have been developed to investigated the mass spectrum of excited
baryon states. Isgur, Karl, and Capstick {\it et al.} described the
baryon resonances as excited states of three constituent quarks
($qqq$) which are confined by a phenomenological confinement
potential and interact through a residual interaction inspired by
one gluon exchange (OGE) \cite{Isgur78,Capstick86}. Glozman and
Riska {\it et al.} proposed a rather different interaction
mechanism. In their model, two quarks interact via Goldstone boson
exchanges (GBE) in addition to a phenomenological confinement
potential, and it is claimed that the flavor-dependent interaction
is responsible for the low mass of the Roper resonance ($N^*(1440)$)
\cite{Glozman96,Glozman00}. So far it is not clear whether the
interactions among the three constituent quarks, which are assumed
to form the baryon resonances, should be described by either OGE or
GBE or a mixture of both \cite{Isgur00,Glozman99}. In chiral
constituent quark models, it is found that some nucleon resonances
are able to be accommodated as baryon-meson dynamically generated
resonances \cite{fhuang05lksk,fhuang05dklksk,fhuang07kbn}. In
Refs.~\cite{fhuang05lksk,fhuang05dklksk}, the $\Lambda K$ and
$\Sigma K$ states have been dynamically investigated in a chiral
SU(3) quark model, and it is shown that a resonance with the same
quantum numbers as the $S_{11}$ nucleon resonances can be
dynamically generated due to the strong $\Sigma K$ attraction. Also
in Ref.~\cite{fhuang07kbn} the ${\bar K}N$ and $\pi\Sigma$
interactions have been dynamically investigated within the extended
chiral SU(3) quark model, and it is found that both the $\pi\Sigma$
and ${\bar K}N$ are bound and the latter appears as a $\pi\Sigma$
resonance in the coupled-channels calculation. This resonance is
referred to $\Lambda(1405)$.

On hadron level, various sophisticated coupled-channel approaches
are formulated for the study of baryon resonances. In the K-matrix
approximation approach \cite{Feuster99,KVI,BonnGatchina}, only
on-shell intermediate states are taken into account when solving the
scattering equation for two-body scattering, which prohibits the
virtual two-body intermediate states. There, all resonances are
treated as genuine resonances and no dynamical poles are reported. The
unitary isobar model is developed by MAID group
\cite{Drechsel:2007if}. It is a variation of the standard K-matrix
approximation approach, and all the resonances are included as
genuine resonances described by the Breit-Wigner forms. In the
chiral unitary approach which includes only the lowest-order
interacting diagrams (i.e. the contact terms) in the scattering
kernel, a completely different picture is delivered and resonances
appear as dynamical effects through the re-scattering. In the baryon
sector, the $N^*(1535)$, $N^*(1650)$, $N^*(1700)$, $\Delta^*(1700)$,
and $\Lambda(1405)$ have been claimed to be dynamically generated
from the interactions of pseudoscalar meson octet or vector meson
octet with nucleon octet or Delta decuplet \cite{KSW95,KL04,SOV05}.
The dynamical coupled-channel hadron-exchange models, capable of a
quantitative description of the meson production processes, have
been developed by J\"ulich group and EBAC group to study the nucleon
resonances \cite{Krehl00,Doring09,RDSHM10,Matsuyama07,Brono07}. In
the J\"ulich model, the Roper ($N^*(1440)$) appears as dynamically
generated resonance and the other resonances like $N^*(1535)$,
$N^*(1650)$ and $\Delta^*(1700)$ are included as genuine resonances
\cite{Krehl00,Doring09,RDSHM10}. In the EBAC model, all resonances
needed by fitting the data are included explicitly and no
dynamically generated resonance is reported
\cite{Matsuyama07,Brono07}.

The situation we have presented so far clearly shows that, the
constituent quark models and the models on hadron level do not give
us a definite picture of the structures of the baryon resonances.
Different models may give us different descriptions for resonances'
structures even though they fit the same set of data, since each
model has its own uncertainties with tunable parameters. Thus it is
still confusing to us whether the baryon resonances should be
described by 3-quark configurations ($qqq$) or 5-quark
configurations ($qqqq{\bar q}$) or baryon-meson dynamically
generated states or a mixture of them.

The study of $\Sigma_c {\bar D}$ and $\Lambda_c {\bar D}$ states is
of particular interest. If there exists a $\Sigma_c {\bar D}$ bound
state or a $\Sigma_c {\bar D}$-$\Lambda_c {\bar D}$ dynamically
generated state, its energy will be around $4.3$ GeV. Unlike the low
energy resonances where the excitation energies, i.e. the energy
differences of nucleon ground state and nucleon resonance states,
are hundreds of MeV which are usually comparable to the $3q$
configuration excitation energy, such a high energy resonance, if it
exists, will have more than $3.3$ GeV excitation energy and thus
will definitely exclude the explanation as three light quark
configuration ($qqq$), and only the description that this state is
dominated by hidden charm five constituent quark configuration
($qqqc{\bar c}$) or $\Sigma_c {\bar D}$ bound state or $\Sigma_c
{\bar D}$-$\Lambda_c {\bar D}$ resonance state or a mixture of them
will be possible.

In Refs.~\cite{wujj101,wujj102}, the interaction between $\Sigma_c
{\bar D}$ and $\Lambda_c {\bar D}$ has been studied within the
framework of the coupled-channel unitary approach. There, a
$\Sigma_c {\bar D}$ bound state is obtained with the energy of
$4.269$ GeV, which is about $52$ MeV below the $\Sigma_c {\bar D}$
threshold. This state is found not to couple to $\Lambda_c {\bar D}$
channel even its energy is about $114$ MeV above the $\Lambda_c
{\bar D}$ threshold. Since the unitary approach used in
Refs.~\cite{wujj101,wujj102} is restricted to the contact term
interaction only by neglecting the momentum-dependent terms, the
study of the $\Sigma_c {\bar D}$ and $\Lambda_c {\bar D}$ state in
other approaches is imperative in order to check the model
dependence and to confirm the possibility of the existence of such a
$\Sigma_c {\bar D}$ bound state.

In the past few years, the chiral SU(3) quark model and its extended
version have shown to be quite reasonable and useful models to
describe the medium-range non-perturbative QCD effect in light
flavor systems. Quite successes have been achieved when these two
models were applied to the studies of the energies of the baryon
ground states, the binding energy of the deuteron, the
nucleon-nucleon ($NN$) and kaon-nucleon ($KN$) scattering phase
shifts of different partial waves, and the hyperon-nucleon ($YN$)
and anti-kaon-nucleon (${\bar K}N$) cross sections
\cite{zyzhang97,lrdai03,fhuang04ctp,fhuang04kn,fhuang04nkdk,fhuang05kne,fhuang08kbn}.
In the chiral SU(3) quark model, the quark-quark interaction
contains OGE, confinement potential, and boson exchanges stemming
from scalar and pseudoscalar nonets. In the extended chiral SU(3)
quark model, the boson exchanges stemming from the vector nonets are
also included, and as a consequence the OGE in largely reduced by
fitting to the energies of the octet and decuplet baryon ground
states. Recently, these two models have also been applied to study
the systems of $N\phi$, $N{\bar\Omega}$, $\Xi{\bar K}$, $\Omega\pi$, $\Omega\omega$,
$\omega\phi$, and $D^0\bar D^{*0}$ {\it et al.}
\cite{fhuang06nphi,dzhang07,wlwang08xikb,wlwang07omepi,wlwang10omfi,wlwang07omeome,liuyr09}.

In this work, we further extend the chiral SU(3) quark model and its
extended version to perform a dynamical coupled-channel study of the
$\Sigma_c \bar D$ and $\Lambda_c \bar D$ states in the framework of
the resonating group method (RGM), a well established method for
studying the interactions among composite particles
\cite{wildermuth77,kamimura77,oka81}. The quark configuration of the
considered system is $(qqc)$-$(q{\bar c})$ with $q$ being the
light-flavor quark $u$ or $d$. We take the interaction between the
light-flavor quark pair $qq$ from our previous works where
the parameters are fixed by a fitting of the energies of octet and
decuplet baryon ground states, the binding energy of deuteron, the
$NN$ scattering phase shifts, and the $YN$ cross sections
\cite{zyzhang97,lrdai03}. The light-heavy quark pair $qc$ or $q{\bar
c}$ and the heavy-heavy quark pair $c{\bar c}$ are considered here
to be interacted via OGE and confinement potential. The only
adjustable parameter is the charm quark mass $m_c$, while the
parameters of OGE and confinement for $qc$, $q{\bar c}$ and $c{\bar
c}$ interactions are fixed by the masses of charmed baryons
$\Sigma_c$, $\Lambda_c$ and charmed mesons $D$, $D^*$ and the
charmonium $J/\psi$, $\eta_c$, and by the stability conditions of
those hadrons. Our results show that the interaction between $\Sigma_c$
and $\bar D$ is attractive, which consequently results in a $\Sigma_c
\bar D$ bound state with the binding energy of about $5-42 $ MeV,
unlike the case of $\Lambda_c \bar D$ state, which has a repulsive
interaction and thus is unbound. The channel coupling effect of
$\Sigma_c \bar D$ and $\Lambda_c \bar D$ is found to be negligible
due to the fact that the gap between the $\Sigma_c \bar D$ and
$\Lambda_c \bar D$ thresholds is relatively large and the $\Sigma_c
\bar D$ and $\Lambda_c \bar D$ transition interaction is weak.

The paper is organized as follows. In the next section the framework
is briefly introduced. The results for the $\Sigma_c \bar D$ and
$\Lambda_c \bar D$ states are shown in Sec.~III, where some
discussion is presented as well. Finally, the summary is given in
Sec.~IV.

\section{Formulation} \label{sec:formulism}

The chiral quark model used in the present work has been widely
described in the literature
\cite{fhuang04kn,fhuang04nkdk,fhuang05kne,fhuang05lksk,fhuang05dklksk,fhuang08kbn},
and we refer the reader to those references for details. Here we
just present the salient features of this model. The total
Hamiltonian is written as
\begin{equation}
H=\sum_{i}T_{i}-T_{G}+\sum_{i,j}V_{ij},
\end{equation}
where $T_i$ is the kinetic energy operator for the $i$th quark, and
$T_G$ the kinetic energy operator for the center-of-mass motion.
$V_{ij}$ represents the interactions between quark-quark or
quark-antiquark,
\begin{equation}
V_{ij}=\left\{ \begin{array}{lll}
V_{ij}^{\rm OGE}+V_{ij}^{\rm conf}+\sum_{\rm M}V_{ij}^{\rm M}, && \left(ij=qq\right) \\[5pt]
V_{ij}^{\rm OGE}+V_{ij}^{\rm conf}, && \left(ij=qQ,q{\bar Q},Q{\bar
Q}\right)
\end{array} \right.
\end{equation}
where $q$ and $Q$ represent light quark $u$ or $d$ and heavy quark $c$, respectively; $V_{ij}^{\rm OGE}$ is the OGE potential,
\begin{eqnarray}
V^{\rm OGE}_{ij} &=& \frac{1}{4}\,g_{i}\,g_{j}\left(\lambda^c_i\cdot\lambda^c_j\right) \Bigg[\frac{1}{r_{ij}} -\frac{\pi}{2} \delta({\bm r}_{ij}) \nonumber \\
&& \times\left(\frac{1}{m^2_{q_i}}+\frac{1}{m^2_{q_j}}+\frac{4}{3}\frac{{\bm \sigma}_i \cdot {\bm \sigma}_j}{m_{q_i}m_{q_j}} \right)\Bigg],
\end{eqnarray}
and $V_{ij}^{\rm conf}$ is the confinement potential which provides
the non-perturbative QCD effect in the long distance,
\begin{eqnarray} \label{eq:conf}
V_{ij}^{\rm conf}=-({\bm \lambda}_{i}^{c}\cdot{\bm
\lambda}_{j}^{c})\left(a_{ij}^{c}r_{ij} +a_{ij}^{c0}\right).
\end{eqnarray}
$V_{ij}^{\rm M}$ represents the effective quark-quark potential
induced by one-boson exchanges, and it is only considered for the
light quark pairs.
%Since the current masses of the heavy quarks are very close to the actual values, their vacuum spontaneous breaking effect turns to unimportant due to the small constituent part. Therefore, the OGE and the confinement potential are enough to describe the main properties in the heavy quark systems. This basic framework is consistent with the QCD inspire model and it is already used in many works \cite{zhanghx08,liuyr09}. The corresponding chiral field forms are,
Generally,
\begin{equation}
V_{ij}^{\rm M} = V_{ij}^{\sigma_a} + V_{ij}^{\pi_a} + V_{ij}^{\rho_a},
\end{equation}
with $V_{ij}^{\sigma_a}$, $V_{ij}^{\pi_a}$ and $V_{ij}^{\rho_a}$ being stemmed from scalar nonets, pseudo-scalar nonets and vector nonets, respectively. Their explicit forms are
\begin{equation}
V^{\sigma_a}({\bm r}_{ij})=-C(g_{ch},m_{\sigma_a},\Lambda) X_1(m_{\sigma_a},\Lambda,r_{ij})
\left(\lambda^a_i\lambda^a_j\right),
\end{equation}
\begin{eqnarray}
V^{\pi_a}({\bm r}_{ij})&=& C(g_{ch},m_{\pi_a},\Lambda) \frac{m^2_{\pi_a}}{12m_{q_i}m_{q_j}} X_2(m_{\pi_a},\Lambda,r_{ij}) \nonumber \\ && \times \left({\bm \sigma}_i\cdot{\bm \sigma}_j\right)
\left(\lambda^a_i\lambda^a_j\right),
\end{eqnarray}
\begin{eqnarray}
V^{\rho_a}({\bm r}_{ij}) &=& C(g_{\rm chv},m_{\rho_a},\Lambda)\Bigg[
X_1(m_{\rho_a},\Lambda,r_{ij}) + \frac{m^2_{\rho_a}}{6m_{q_i}m_{q_j}} \nonumber \\
&& \times \left(1+\frac{f_{\rm chv}}{g_{\rm chv}} \frac{m_{q_i}+m_{q_j}}{M_N}+\frac{f^2_{\rm chv}}{g^2_{\rm chv}}  \frac{m_{q_i}m_{q_j}}{M^2_N}\right)   \nonumber \\
&& \times \,  X_2(m_{\rho_a},\Lambda,r_{ij}) \, ({\bm \sigma}_i\cdot{\bm \sigma}_j)
\Bigg] \left(\lambda^a_i\lambda^a_j\right),
\end{eqnarray}
where
\begin{eqnarray}
C(g_{\rm ch},m,\Lambda) &=& \frac{g^2_{\rm ch}}{4\pi} \frac{\Lambda^2}{\Lambda^2-m^2} m, \\
\label{x1mlr} X_1(m,\Lambda,r) &=& Y(mr)-\frac{\Lambda}{m} Y(\Lambda r), \\
X_2(m,\Lambda,r) &=& Y(mr)-\left(\frac{\Lambda}{m}\right)^3 Y(\Lambda r), \\
Y(x) &=& \frac{1}{x}e^{-x},
\end{eqnarray}
with $m_{\sigma_a}$ being the mass of the scalar meson, $m_{\pi_a}$
the mass of the pseudoscalar meson and $m_{\rho_a}$ the mass of the
vector meson. $m_{q_i}$ is the constituent quark mass of the $i$th
quark. $g_{\rm ch}$ is the coupling constant for the scalar and
pseudoscalar nonets, and $g_{\rm chv}$ and $f_{\rm chv}$ the
coupling constants for vector coupling and tensor coupling of vector
nonets.

In this work, we take the parameters for light-flavor quark system from our previous works
\cite{fhuang05dklksk,wlwang08xikb,wlwang07omepi}, which gave a
satisfactory description for the energies of the octet and decuplet
baryon ground state, the binding energy of the deuteron, the $NN$
scattering phase shifts, and the $NY$ cross sections. The main
procedure for determination of those parameters is the following.
The initial input parameters, i.e. the harmonic-oscillator width
parameter $b_u$ and the up (down) quark mass $m_{u(d)}$, are taken
to be the usual values: $b_u=0.5$ fm for the chiral SU(3) quark
model and $0.45$ fm for the extended chiral SU(3) quark model,
$m_{u(d)}=313$ MeV. The coupling constant for scalar and
pseudoscalar chiral field coupling, $g_{\rm ch}$, is fixed by the
relation
\begin{eqnarray}
\frac{g^{2}_{ch}}{4\pi} = \left( \frac{3}{5} \right)^{2} \frac{g^{2}_{NN\pi}}{4\pi} \frac{m^{2}_{u}}{M^{2}_{N}},
\end{eqnarray}
with the empirical value $g^{2}_{NN\pi}/4\pi=13.67$. For the vector
meson field coupling, we consider three different cases. In model I,
the coupling between vector meson field and quark field is not
considered at all, which means $g_{\rm chv}=0$. Then in model II and
III, the coupling constant for vector coupling is taken to be
$g_{\rm chv}=2.351$ and $1.973$, respectively, and the ratio for the
tensor coupling and vector coupling is taken to be $0$ and $2/3$,
respectively. The masses of the mesons are taken to be the
experimental values, except for the $\sigma$ meson. The $m_\sigma$
is obtained by fitting the binding energy of the deuteron. The
cutoff radius $\Lambda^{-1}$ is taken to be the value close to the
chiral symmetry breaking scale \cite{ito90,amk91,abu91,emh91}. The
OGE coupling constants and the strengths of the confinement
potential are fitted by the baryon masses and their
stability conditions. % After the parameters of chiral fields are
%fixed, the one-gluon-exchange coupling constants $g_{u}$ is
%determined by the mass splits between $N$ and $\Delta$. The
%confinement strengths $a^{c}_{uu}$ and the zero-point energies
%$a^{c0}_{uu}$ are fixed by the mass and the stability condition of
%$N$.

Note that in light-flavor quark systems, the confinement potential
is found to give negligible contributions between two color-singlet
hadron clusters
\cite{fhuang04kn,fhuang04nkdk,fhuang05kne,fhuang05lksk}. Therefore
different forms of confinement potential (linear or quadratic) does
not make any visible influence on the theoretical results in the
light-flavor quark systems. In the present work we adopt a color
linear confinement potential. The results from a calculation by
using the color quadratic confinement potential are discussed as
well. Of course the $NN$ scattering phase shifts and the $NY$ cross
sections are always well described irrespective of confinement forms
due to the negligibility of the contributions of any confinement to
these systems.

The additional parameters needed in the present work are those
associated with charm quark. The only one adjustable parameter is
the charm quark mass $m_c$. Here we take three typical values,
$m_c=1.43$ GeV \cite{zhanghx08}, $1.55$ GeV \cite{vijande04} and
$1.87$ GeV \cite{bsb93}, to test the dependence of our results on
$m_c$. The other parameters we need are the coupling constant of OGE
and confinement strengths for light quark and heavy quark pair, $qc$
and $q{\bar c}$, and for heavy quark pair, $c{\bar c}$. They are
fixed by a fitting to the masses and stability conditions of the
charmed baryons $\Sigma_c$, $\Lambda_c$ and charmed mesons $D$,
$D^*$ and the charmonium $J/\psi$, $\eta_c$. The values of those
parameters are listed in Table~\ref{para}. The corresponding masses
of $\Sigma_c$, $\Lambda_c$, $D$, $D^*$, $J/\psi$ and $\eta_c$
obtained with $m_c=1.55$ GeV are shown in Table~\ref{mass}.
There, Model I refers to the model where the coupling for vector
nonets is not considered. Models II and III refer to the models
where the coupling for vector nonets is included while the ratio for
tensor coupling and vector coupling $f_{\rm chv}/g_{\rm chv}$ is
taken to be $0$ and $2/3$, respectively.

\begin{table}[tb]
\caption{\label{para}Model parameters. Model I refers to the model
where the coupling for vector nonets is not considered. Models II
and III refer to the models where the coupling for vector nonets is
included while the ratio for tensor coupling and vector coupling
$f_{\rm chv}/g_{\rm chv}$ is taken to be $0$ and $2/3$,
respectively. }
\renewcommand{\arraystretch}{1.3}
\begin{tabular*}{0.49\textwidth}{@{\extracolsep\fill}ccccccccc}
\hline\hline
  & $m_c$ & $g_c$ & $a^c_{uu}$ & $a^c_{uc}$ & $a^c_{cc}$ & $a^{c0}_{uu}$ & $a^{c0}_{uc}$ & $a^{c0}_{cc}$ \\
          & (GeV) & & (fm$^{-2}$)& (fm$^{-2}$) & (fm$^{-2}$) & (fm$^{-1}$)& (fm$^{-1}$) & (fm$^{-1}$) \\
 \hline
 I  & $1.43$ & $0.35$ & $0.44$ & $1.07$ & $1.74$ & $-0.38$ & $-0.74$ & $-0.73$  \\
     & $1.55$ & $0.37$ & $0.44$ & $1.08$ & $1.77$ & $-0.38$ & $-0.85$ & $-0.93$  \\
     & $1.87$ & $0.43$ & $0.44$ & $1.10$ & $1.81$ & $-0.38$ & $-1.14$ & $-1.44$  \\
 II  & $1.43$ & $0.77$ & $0.41$ & $1.70$ & $1.83$ & $-0.53$ & $-1.15$ & $-0.34$  \\
     & $1.55$ & $0.82$ & $0.41$ & $1.72$ & $1.68$ & $-0.53$ & $-1.27$ & $-0.40$  \\
     & $1.87$ & $0.94$ & $0.41$ & $1.76$ & $1.04$ & $-0.53$ & $-1.57$ & $-0.47$  \\
 III & $1.43$ & $0.57$ & $0.37$ & $1.68$ & $2.19$ & $-0.46$ & $-1.14$ & $-0.71$  \\
     & $1.55$ & $0.60$ & $0.37$ & $1.69$ & $2.16$ & $-0.46$ & $-1.25$ & $-0.85$  \\
     & $1.87$ & $0.69$ & $0.37$ & $1.74$ & $1.94$ & $-0.46$ & $-1.55$ & $-1.17$  \\
\hline\hline
\end{tabular*}
\end{table}

\begin{table}[tb]
\caption{\label{mass} The masses (in GeV) of $\Sigma_c$, $\Lambda_c$, $D$,
$D^*$, $J/\psi$ and $\eta_c$ obtained from models I, II and III, respectively, with $m_c$ being taken as $1.55$ GeV. Experimental values are taken from PDG \cite{PDG2010}.}
\renewcommand{\arraystretch}{1.3}
\begin{tabular*}{0.49\textwidth}{@{\extracolsep\fill}ccccccc}
\hline\hline
  & $\Sigma_c$ & $\Lambda_c$ & $D$ & $D^*$ & $J/\psi$ & $\eta_c$ \\
\hline
 Exp.   & $2.452$ & $2.286$ & $1.869$ & $2.007$ & $3.097$ & $2.980$ \\
 I    & $2.436$ & $2.269$ & $1.883$ & $1.947$ & $3.052$ & $3.024$ \\
II  & $2.450$ & $2.283$ & $1.869$ & $1.932$ & $3.129$ & $2.946$ \\
III    & $2.450$ & $2.283$ & $1.869$ & $1.932$ & $3.087$ & $2.989$ \\
\hline\hline
\end{tabular*}
\end{table}

With all parameters determined, the $\Sigma_c \bar D$ and $\Lambda_c
\bar D$ systems can be dynamically studied in the frame work of the
RGM, where the wave function of the five-quark system is of the
following form:
\begin{eqnarray} \label{eq:rgm}
\Psi=\sum_\beta {\cal A}\left\{\left[{\hat \phi}_A(\bm \xi_1,\bm \xi_2) {\hat \phi}_B(\bm \xi_3)\right]_\beta \chi_{\beta}({\bm R}_{AB})\right\}.
\end{eqnarray}
Here ${\bm \xi}_1$ and ${\bm \xi}_2$ are the internal coordinates
for the cluster $A$ ($\Lambda_c$ or $\Sigma_c$), and ${\bm \xi}_3$
the internal coordinate for the cluster $B$ ($\bar D$). ${\bm
R}_{AB}\equiv {\bm R}_A-{\bm R}_B$ is the relative coordinate
between the two clusters, $A$ and $B$, and $\beta\equiv
(A,B,I,S,L,J)$ specifies the hadron species ($A$, $B$) and quantum
numbers of the baryon-meson channel. The ${\hat \phi}_A$ and ${\hat
\phi}_B$ are the internal cluster wave functions of $A$ and $B$, and
$\chi_\beta ({\bm R}_{AB})$ the relative wave function of the two
clusters. The symbol $\cal A$ is the anti-symmetrizing operator
defined as
\begin{equation}
{\cal A}\equiv{1-\sum_{i \in A}P_{i4}}\equiv{1-3P_{34}}.
\end{equation}
Substituting $\Psi$ into the projection equation
\begin{equation}
\langle \delta\Psi|(H-E)|\Psi \rangle=0,
\end{equation}
we obtain the coupled integro-differential equation for the relative
function $\chi_\beta$ as
\begin{equation}\label{crgm}
\sum_{\beta'}\int \left[{\cal H}_{\beta\beta'}(\bm R, \bm R')-E{\cal N}_{\beta\beta'}(\bm R, \bm R')\right]\chi_{\beta'}(\bm R') \,{\rm d}\bm R' =0,
\end{equation}
where the Hamiltonian kernel $\cal H$ and normalization kernel $\cal N$ can, respectively, be calculated by
\begin{eqnarray}
\left\{
       \begin{array}{c}
          {\cal H}_{\beta\beta'}(\bm R, \bm R')\\
          {\cal N}_{\beta\beta'}(\bm R, \bm R')
       \end{array}
\right\} =\Bigg<[{\hat \phi}_A(\bm \xi_1,\bm \xi_2 ) {\hat
\phi}_B(\bm \xi_3)]_\beta\delta(\bm R-{\bm R}_{AB})  \nonumber \\
\left|
\left\{
\begin{array}{c}
          H \\
          1
       \end{array}
\right\}
\right| {\cal A}\left[[{\hat \phi}_A(\bm \xi_1,\bm \xi_2) {\hat
\phi}_B(\bm \xi_3)]_{\beta'}\delta(\bm R'-{\bm
R}_{AB})\right]\Bigg>. \nonumber \\
\end{eqnarray}

Equation (\ref{crgm}) is the so-called coupled-channel RGM equation.
Expanding unknown $\chi_\beta ({\bm R}_{AB})$ by employing
well-defined basis wave functions, such as Gaussian functions, one
can solve the coupled-channel RGM equation for a bound-state problem
or a scattering one to obtain the binding energy or scattering $S$
matrix elements for the two-cluster systems. The details of solving
the RGM equation can be found in
Refs.~\cite{wildermuth77,kamimura77,oka81}.

\section{Results and discussions}

As mentioned in the Introduction, the structures of the nucleon
resonances below $2$ GeV are not clear so far. Different models may
give us different pictures even they fit the same set of data, since
each model has its own uncertainties which are usually approximated
by fitting parameters. It is still a challenging task for hadron
physicist whether the low energy baryon resonances should be
described by three constituent quark configuration ($qqq$) or five
constituent quark configuration ($qqqq{\bar q}$) or baryon-meson
dynamically generated states or a mixture of them. The $\Sigma_c
\bar D$ and $\Lambda_c \bar D$ states are of particular interest
simply because if there exists a $\Sigma_c \bar D$ bound state or a
$\Sigma_c \bar D$-$\Lambda_c \bar D$ dynamically generated
resonance, its energy will be around $4.3$ GeV and the explanation
of such a high energy state as three constituent quark configuration
($qqq$) will be definitely excluded while only the description that
this state is dominated by hidden charm five constituent quark
configuration ($qqqc{\bar c}$) or $\Sigma_c \bar D$-$\Lambda_c \bar
D$ baryon-meson state or a mixture of them will be possible.
Thus the system of $\Sigma_c \bar D$-$\Lambda_c \bar D$ will be a good place to test whether we could have a nucleon resonance whose configuration is dominated by at least five quarks.

Here we perform a dynamical investigation of the $\Sigma_c \bar D$
and $\Lambda_c \bar D$ states with isospin $I=1/2$ and spin $S=1/2$
by solving the RGM equation (Eq.~(\ref{crgm})) in our chiral quark
models as depicted in Sec.~\ref{sec:formulism}. Our purpose is to
understand the interaction properties of the $\Sigma_c \bar D$ and
$\Lambda_c \bar D$ states and to see whether there exists a
$\Sigma_c \bar D$ bound state or a $\Sigma_c \bar D$-$\Lambda_c \bar
D$ dynamically generated resonance within our chiral quark models.

\begin{figure}
\centering\includegraphics[width=0.49\textwidth,clip]{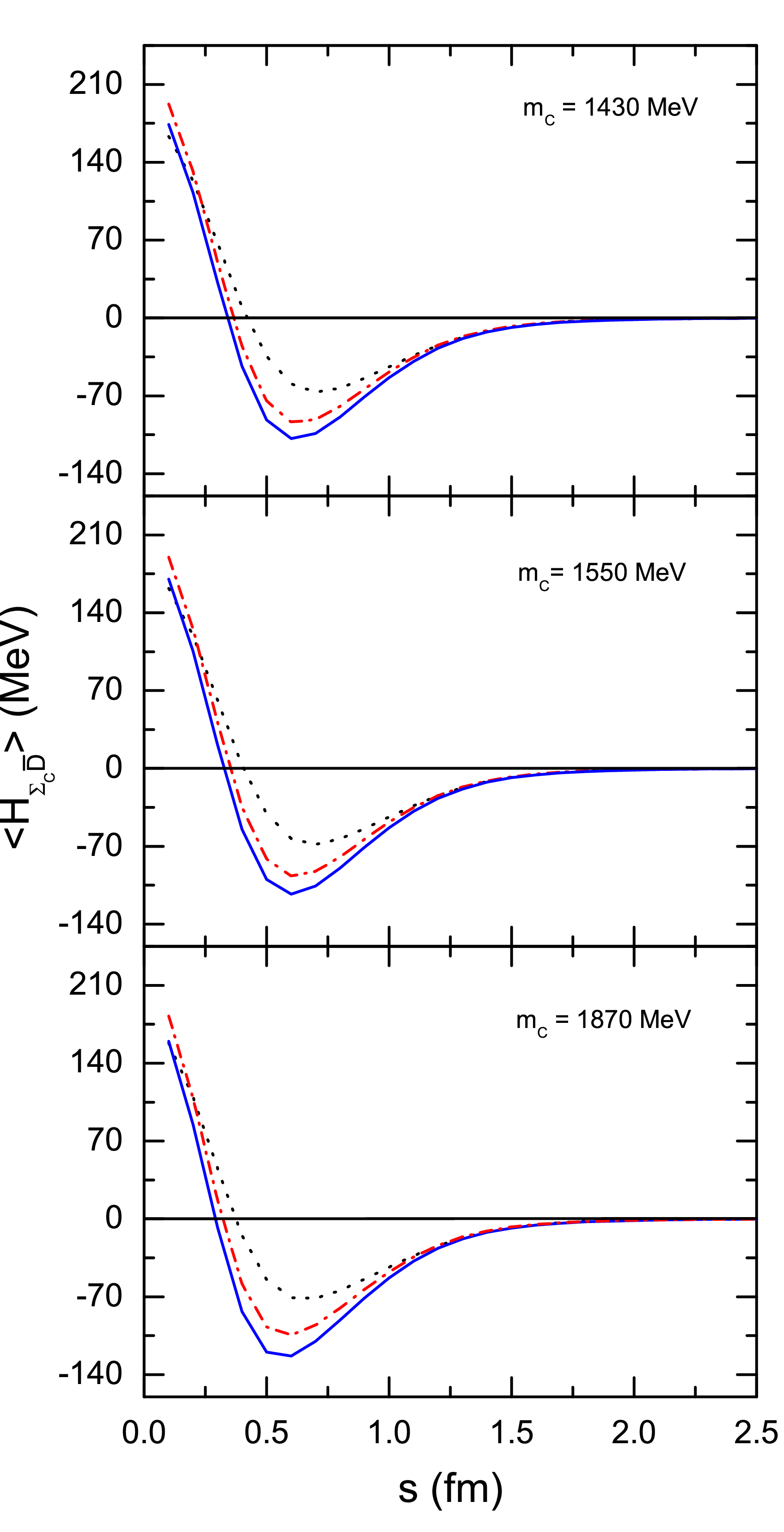}
\caption{\label{HS} The GCM matrix elements of the Hamiltonian for $\Sigma_c \bar D$ system. The dotted, solid and dash-dotted lines represent the results obtained in models I, II and III, respectively.}
\end{figure}

\begin{figure}
\centering\includegraphics[width=0.49\textwidth,clip]{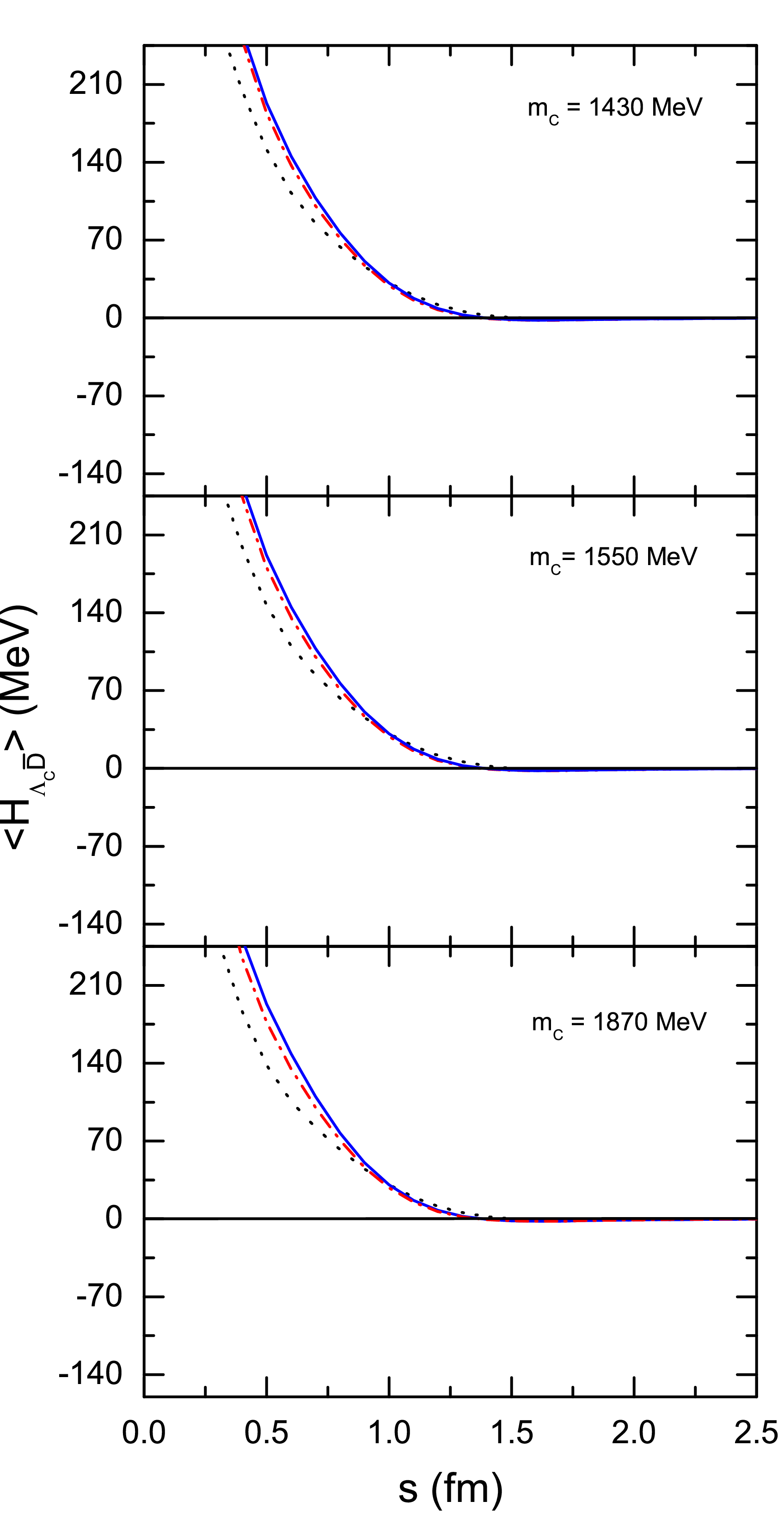}
\caption{\label{HL} The GCM matrix elements of the Hamiltonian for $\Lambda_c \bar D$ system. The dotted, solid and dash-dotted lines represent the results obtained in models I, II and III, respectively.}
\end{figure}

Figure \ref{HS} shows the diagonal matrix elements of the
Hamiltonian for the $\Sigma_c {\bar D}$ system in the generator
coordinate method (GCM) \cite{wildermuth77} calculation, which can
be regarded as the effective Hamiltonian of two color-singlet
clusters $\Sigma_c$ and $\bar D$ qualitatively. In Fig.~\ref{HS},
$H_{\Sigma_c \bar D}$ includes the kinetic energy of $\Sigma_c \bar
D$ relative motion and the effective potential between $\Sigma_c$
and $\bar D$, and $s$ denotes the generator coordinate which can
qualitatively describe the distance between the two clusters
$\Sigma_c$ and $\bar D$. From Fig.~\ref{HS}, one sees that
$\Sigma_c$ and $\bar D$ are attractive to each other in the medium
range for all those three values of charm quark mass $m_c=1.43$ GeV,
$1.55$ GeV and $1.87$ GeV and all those three models I, II and III
(see Sec.~\ref{sec:formulism} for details of these three models).
Our further analysis demonstrates that in model I the attraction
between $\Sigma_c$ and $\bar D$ is dominated by $\sigma$ exchange
and the color magnetic force of OGE; the latter exists between the
two color-singlet clusters $\Sigma_c$ and $\bar D$ because of the
anti-symmetrizing (Eq.~(\ref{eq:rgm})) of the four constituent
quarks in $\Sigma_c\bar D$ required by the general Pauli principle.
In models II and III, the OGE among light-flavor quarks are largely
reduced by vector-meson exchanges and the $\Sigma_c{\bar D}$
attraction is found to be dominated by $\sigma$ and $\rho$
exchanges.

Inspired by the moderately large $\Sigma_c \bar D$ attraction, we
have solved the RGM equation for a bound state problem to see
whether there is a $\Sigma_c \bar D$ bound state or not. Our results
are listed in Table \ref{ener}, where the first and second columns
denote the model and the charm quark mass, respectively, and the
third column shows the corresponding binding energy for each set of
parameters. One sees that the $\Sigma_c \bar D$ is really bound
independent of the types of the models and the values of the charm
quark mass we use. The binding energy is around $9-42$ MeV in
various models, which corresponding to an energy of $4.279-4.312$
GeV for the $\Sigma_c \bar D$ bound state (the $\Sigma_c {\bar D}$
threshold is $4.321$ GeV).

\begin{table}[tb]
\caption{\label{ener}The binding energy of $\Sigma_c \bar D$ (in MeV) in models I, II and III, respectively.}
\renewcommand{\arraystretch}{1.3}
\begin{tabular*}{0.49\textwidth}{@{\extracolsep\fill}cccc}
\hline\hline
  & $m_c$ (GeV) & $r$ confinement & $r^2$ confinement \\
\hline
  I  & $1.43$ & $~9.3$  & $~4.5$   \\
     & $1.55$ & $10.9$  & $~6.4$   \\
     & $1.87$ & $15.3$  & $11.0$   \\
 II  & $1.43$ & $28.3$  & $~9.3$   \\
     & $1.55$ & $31.8$  & $10.3$   \\
     & $1.87$ & $41.6$  & $10.0$   \\
 III & $1.43$ & $19.7$  & $~7.3$   \\
     & $1.55$ & $22.2$  & $~8.9$   \\
     & $1.87$ & $28.6$  & $11.3$   \\
\hline\hline
\end{tabular*}
\end{table}

Here we'd like to discuss the dependence of our results on the
phenomenology confinement potential. In light-flavor quark systems,
the SU(3) flavor symmetry is approximately respected and thus the
confinement potential is found to give negligible contributions
between two color-singlet hadron clusters
\cite{fhuang04kn,fhuang04nkdk,fhuang05kne,fhuang05lksk}. As far as
the charm quark is included, the SU(4) flavor symmetry is strongly
violated since the charm quark mass is much bigger than that of
light-flavor quark. The consequence of this flavor symmetry
violation is that the contribution of the confinement potential to
the interaction between two hadron clusters may not be negligible.
In the present work, we check the dependence of our results on the
forms of the confinement potential by replacing the linear
confinement (Eq.~(\ref{eq:conf})) with the quadratic one,
\begin{eqnarray} \label{eq:conf2}
V_{ij}^{\rm conf}=-({\bm \lambda}_{i}^{c}\cdot{\bm
\lambda}_{j}^{c})\left(a_{ij}^{c}r^2_{ij} +a_{ij}^{c0}\right),
\end{eqnarray}
with the parameters being fitted by using the same procedure as
given in the previous section. With the quadratic confinement
Eq.~(\ref{eq:conf2}), we re-solve the RGM equation for $\Sigma_c
{\bar D}$ bound state problem, and the results are shown in the
fourth column of Table \ref{ener}. One sees that the $\Sigma_c {\bar
D}$ is still bound in various models and the binding energy is
around $5-11$ MeV which is a little smaller than that for the linear
confinement. The corresponding energy of $\Sigma_c {\bar D}$ bound
state is $4.310-4.316$ GeV.

We have also studied the $\Lambda_c {\bar D}$ system. Figure
\ref{HL} shows the diagonal matrix elements of the Hamiltonian for
the $\Lambda_c {\bar D}$ system in the GCM calculation, which can be
regarded as the effective Hamiltonian of two color-singlet clusters
$\Lambda_c$ and $\bar D$ qualitatively. One sees that unlike the
$\Sigma_c {\bar D}$ system which is attractive in the medium range,
the $\Lambda_c{\bar D}$ system is strongly repulsive for all those
three models and all those three values of charm quark mass. No
$\Lambda_c{\bar D}$ bound state will be found as a matter of course
due to this repulsion.

Is there a $\Sigma_c {\bar D}$-$\Lambda_c{\bar D}$ resonance in the
coupled-channel study? In Refs.~\cite{fhuang05lksk,fhuang05dklksk},
we have dynamically investigated the $\Sigma K$ and $\Lambda K$
systems by using RGM in our chiral quark model. There, it is found
that the $\Sigma K$ interaction is attractive and a $\Sigma K$ bound
state can be formed as a consequence with the binding energy of
about $17-44$ MeV, while the $\Lambda K$ is repulsive and unbound.
In the coupled-channel calculation, a $\Sigma K$-$\Lambda K$
dynamically generated resonance is obtained which is located between
the thresholds of $\Sigma K$ and $\Lambda K$ and has the quantum
numbers the same as those for nucleon $S_{11}$ resonances.
Analogically, one may expect a $\Sigma_c {\bar D}$-$\Lambda_c{\bar
D}$ dynamically generated resonance in the coupled-channel
calculation since $\Sigma_c {\bar D}$ is also attractive and bound
just like $\Sigma K$. But actually, the coupled-channel effect of
$\Sigma_c {\bar D}$ and $\Lambda_c{\bar D}$ is found to be
negligible, and no $\Sigma_c {\bar D}$-$\Lambda_c{\bar D}$ resonance
is found in our coupled-channel calculation. This is because the gap
of the $\Sigma_c {\bar D}$ and $\Lambda_c{\bar D}$ thresholds, $166$
MeV, is comparatively big and the transition matrix elements between
$\Sigma_c {\bar D}$ and $\Lambda_c{\bar D}$ are too weak, contrary
to the case of $\Sigma K$-$\Lambda K$ system, where the gap of two
channel thresholds is only $78$ MeV and the transition matrix
elements between $\Sigma K$ and $\Lambda K$ are relatively large.

In brief, we obtain a $\Sigma_c {\bar D}$ bound state in our model
with the energy of about $4.279-4.316$ MeV, and the effect from
$\Lambda_c {\bar D}$ channel to this state is negligible. In
Refs.~\cite{wujj101,wujj102}, the $\Sigma_c {\bar D}$ and $\Lambda_c
{\bar D}$ states have been studied on hadron level within the
framework of the coupled-channel unitary approach. There, a
$\Sigma_c {\bar D}$ bound state is also found with the energy of
about $4.240-4.291$ GeV, and this state does not couple to
$\Lambda_c {\bar D}$ channel. Although the binding energy given by
Refs.~\cite{wujj101,wujj102} is bigger than what we get from the
present work, it makes sense that the results from different
theoretical approaches are qualitatively similar. Note that the
$\Sigma_c {\bar D}$ bound state, if it exists, cannot be
accommodated in three light flavor quark configuration ($qqq$),
unlike the nucleon resonances below $2$ GeV. Whether it can be
explained as hidden charm five constituent quark configuration
($qqqc{\bar c}$) or not needs further detailed scrutiny.
Investigations from other approaches and experiments are needed to
further confirm the existence of this state and to pin down its
structure and mass. Since its mass is above the $\eta_c N$ and
$J/\psi N$ thresholds, it is much easier for their experimental
searches~\cite{wujj101,wujj102}, compared with those baryons with
hidden charms below the $\eta_cN$ threshold proposed by other
approaches~\cite{Brodsky1}.

\section{Summary}

In this work, we perform a dynamical coupled-channel study of
$\Sigma_c \bar D$ and $\Lambda_c \bar D$ states by solving the RGM
equation in the framework of a chiral quark model. The model
parameters for light-flavor quarks are taken from our previous work
\cite{fhuang05dklksk}, which gave a satisfactory description of the
energies of the octet and decuplet baryon ground states, the binding
energy of the deuteron, the $NN$ scattering phase shifts, and the
$NY$ cross sections. The parameters associated with charm quark are
determined by fitting the energies and the stability conditions of
$\Sigma_c$, $\Lambda_c$, $D$, $D^*$, $J/\psi$ and $\eta_c$. Our
results show that the $\Sigma_c$ and $\bar D$ interaction is
attractive and a $\Sigma_c {\bar D}$ bound state can be formed as a
consequence with the energy of about $4.279-4.316$ GeV, while the
$\Lambda_c \bar D$ is repulsive and unbound. The channel-coupling
effects between $\Sigma_c \bar D$ and $\Lambda_c \bar D$ is
negligible due to the large mass difference between the $\Sigma_c
\bar D$ and $\Lambda_c \bar D$ thresholds and the small off-diagonal
matrix elements of $\Sigma_c \bar D$ and $\Lambda_c \bar D$. This
$\Sigma_c {\bar D}$ bound state, if it really exists, cannot be
accommodated in three light flavor quark configuration ($qqq$).
Further investigations from other approaches and experiments are
needed to confirm the existence of this state and to pin down its
structure and mass.

\begin{acknowledgments}
This work was supported in part by the National Natural Science
Foundation of China grant Nos. 10875133, 10821063 and Ministry of
Science and Technology of China (2009 CB 825200) and China
Postdoctoral Science Foundation (No. 20100480468). F.H. is grateful
to the support by COSY FFE grant No. 41788390 (COSY-058).
\end{acknowledgments}


\begin{thebibliography}{99}
\bibitem{Isgur78}N. Isgur and G. Karl, Phys. Rev. D \textbf{18}, 4187 (1978).
\bibitem{Capstick86}S. Capstick and N. Isgur, Phys. Rev. D \textbf{34}, 2809 (1986).
\bibitem{Glozman96}L.Ya. Glozman and D.O. Riska, Phys. Rept. \textbf{268}, 263 (1996).
\bibitem{Glozman00}L.Ya. Glozman, Nucl. Phys. A \textbf{663}, 103c (2000).
\bibitem{Isgur00}N. Isgur, Phys. Rev. D \textbf{62}, 054026 (2000).
\bibitem{Glozman99}L.Ya. Glozman, arXiv: nucl-th/9909021.
\bibitem{fhuang05lksk}F. Huang, D. Zhang, Z.Y. Zhang, and Y.W. Yu, Phys. Rev. C \textbf{71}, 064001 (2005).
\bibitem{fhuang05dklksk}F. Huang and Z.Y. Zhang, Phys. Rev. C \textbf{72}, 068201 (2005).
\bibitem{fhuang07kbn}F. Huang, W.L. Wang, Z.Y. Zhang, and Y.W. Yu, Phys. Rev. C \textbf{76}, 018201 (2007).
\bibitem{Feuster99}V. Shklyar, G. Penner, and U. Mosel, Eur. Phys. J. A\textbf{21}, 445 (2004).
\bibitem{KVI}A. Usov and O. Scholten, Phys. Rev. C \textbf{72}, 025205 (2005).
\bibitem{BonnGatchina}A.V. Sarantsev, V.A. Nikonov, A.V. Anisovich, E. Klempt, and U. Thoma, Eur. Phys. J. A         \textbf{25}, 441 (2005).
\bibitem{Drechsel:2007if}D. Drechsel, S.S. Kamalov, and L. Tiator, Eur. Phys. J. A \textbf{34}, 69 (2007).
\bibitem{KSW95}N. Kaiser, P.B. Siegel, and W. Weise, Phys. Lett. B \textbf{362}, 23 (1995).
\bibitem{KL04}E.E. Kolomeitsev and M.F.M. Lutz, Phys. Lett. B \textbf{585}, 243 (2004).
\bibitem{SOV05}S. Sarkar, E. Oset, and M.J. Vincente Vacas, Nucl. Phys. A \textbf{750}, 294 (2005).
\bibitem{Krehl00}O. Krehl, C. Hanhart, S. Krewald, and J. Speth, Phys. Rev. C \textbf{62}, 025207 (2000).
\bibitem{Doring09}M. D\"oring, C. Hanhart, F. Huang, S. Krewald, and U.-G. Mei\ss ner, Nucl. Phys. A \textbf{829}, 170 (2009).
\bibitem{RDSHM10}M. D\"oring, C. Hanhart, F. Huang, S. Krewald, U.-G. Mei\ss ner, and D. R\"onchen, arXiv: 1009.3781 [nucl-th].
\bibitem{Matsuyama07}A. Matsuyama, T. Sato, and T.-S.H. Lee, Phys. Rept. \textbf{439}, 193 (2007).
\bibitem{Brono07}B. Juli\'a-D\'iaz, T.-S.H. Lee, A. Matsuyama, and T. Sato, Phys. Rev. C \textbf{76}, 065201 (2007).
\bibitem{wujj101}J.J. Wu, R. Molina, E. Oset, and B.S. Zou, Phys. Rev. Lett. \textbf{105}, 232001 (2010).
\bibitem{wujj102}J.J. Wu, R. Molina, E. Oset, and B.S. Zou, arXiv: 1011.2399 [nucl-th].
\bibitem{zyzhang97}Z.Y. Zhang, Y.W. Yu, P.N. Shen, L.R. Dai, A. Faessler, and U. Straub, Nucl. Phys. A \textbf{625}, 59 (1997).
\bibitem{lrdai03}L.R. Dai, Z.Y. Zhang, Y.W. Yu, and P. Wang, Nucl. Phys. A \textbf{727}, 321 (2003).
\bibitem{fhuang04ctp}F. Huang, Z.Y. Zhang, and Y.W. Yu, Commun. Theor. Phys. \textbf{42}, 577 (2004).
\bibitem{fhuang04kn}F. Huang, Z.Y. Zhang, and Y.W. Yu, Phys. Rev. C \textbf{70}, 044004 (2004).
\bibitem{fhuang04nkdk}F. Huang and Z.Y. Zhang, Phys. Rev. C \textbf{70}, 064004 (2004).
\bibitem{fhuang05kne}F. Huang and Z.Y. Zhang, Phys. Rev. C \textbf{72}, 024003 (2005).
\bibitem{fhuang08kbn}F. Huang, W.L. Wang, and Z.Y. Zhang, Int. J. Mod. Phys. A \textbf{23}, 3057 (2008).
\bibitem{fhuang06nphi}F. Huang, Z.Y. Zhang, and Y.W. Yu, Phys. Rev. C \textbf{73}, 025207 (2006).
\bibitem{dzhang07}D. Zhang, F. Huang, L.R. Dai, Y.W. Yu, and Z.Y. Zhang, Phys. Rev. C \textbf{75}, 024001 (2007).
\bibitem{wlwang08xikb}W.L. Wang, F. Huang, Z.Y. Zhang, and F. Liu, J. Phys. G \textbf{35}, 085003 (2008).
\bibitem{wlwang07omepi}W.L. Wang, F. Huang, Z.Y. Zhang, Y.W. Yu, and F. Liu, Eur. Phys. J. A \textbf{32}, 293-297 (2007).
\bibitem{wlwang10omfi}W.L. Wang, F. Huang, Z.Y. Zhang, and F. Liu, Mod. Phys. Lett. A \textbf{25} 1325 (2010).
\bibitem{wlwang07omeome}W.L. Wang, F. Huang, Z.Y. Zhang, Y.W. Yu, and F. Liu, Commun. Theor. Phys. \textbf{48}, 695 (2007).
\bibitem{liuyr09}Y.R. Liu and Z.Y. Zhang, Phys. Rev. C {\bf 79}, 035206 (2009); ibid. Phys. Rev. C {\bf 80}, 015208 (2009).
\bibitem{wildermuth77}K. Wildermuth and Y.C. Tang, {\it A Unified Theory of the Nucleus}, Vieweg, Braunschweig (1977).
\bibitem{kamimura77}M. Kamimura, Suppl. Prog. Theor. Phys. \textbf{62}, 236 (1977).
\bibitem{oka81}M. Oka and K. Yazaki, Prog. Theor. Phys. \textbf{66}, 556 (1981).
\bibitem{ito90}I.T. Obukhovsky and A.M. Kusainov, Phys. Lett. B \textbf{238}, 142 (1990).
\bibitem{amk91}A.M. Kusainov, V.G. Neudatchin, and I.T. Obukhovsky, Phys. Rev. C \textbf{44}, 2343 (1991).
\bibitem{abu91}A. Buchmann, E. Fernandez, and K. Yazaki, Phys. Lett. B \textbf{269}, 35 (1991).
\bibitem{emh91}E.M. Henley and G.A. Miller, Phys. Lett. B \textbf{251}, 453 (1991).
\bibitem{zhanghx08}H.X. Zhang, W.L. Wang, Y.B. Dai, and Z.Y. Zhang, Commun. Theor. Phys. \textbf{49}, 414 (2008).
\bibitem{vijande04}J. Vijande, H. Garcilazo, A. Valcarce, and F. Fernandez, Phys. Rev. D \textbf{70}, 054022 (2004).
\bibitem{bsb93}B. Silvestre-Brac and C. Semay, Z. Phys. C \textbf{57}, 273 (1993).
\bibitem{PDG2010}K. Nakamura {\it et al.} (Particle Data Group), J. Phys. G \textbf{37}, 075021 (2010).
\bibitem{Brodsky1}S.J. Brodsky, I.A. Schmidt, and G.F. de Teramond, Phys. Rev. Lett. {\bf 64}, 1011 (1990); C. Gobbi, D.O. Riska, and N.N. Scoccola, Phys. Lett. B {\bf 296}, 166 (1992); J. Hofmann and M. Lutz, Nucl. Phys. A {\bf 763}, 90 (2005).
\end{thebibliography}
\end{document}